\documentclass[11pt, twocolumn]{aastex631}

\usepackage{graphics}

\begin{document}

\title{ Fast Multipole Method for Gravitational Lensing. Application to High Magnification Quasar Microlensing.}

\correspondingauthor{J. Jim\'enez-Vicente}

\email{jjimenez@ugr.es}

\author[0000-0002-0786-7307]{J. Jim\'enez-Vicente}
\affiliation{Departamento de F\'{\i}sica Te\'orica y del Cosmos. Universidad de Granada. Campus de Fuentenueva,
  18071 Granada, Spain}
\affiliation{Instituto Carlos I de F\'{\i}sica Te\'orica y Computacional. Universidad de Granada.
  18071 Granada, Spain}
\author{E. Mediavilla}
\affiliation{Instituto de Astrof\'{\i}sica de Canarias. V\'{\i}a L\'actea S/N, La Laguna 38200, Tenerife, Spain}
\affiliation{Departamento de Astrof\'{\i}sica. Universidad de la Laguna. La
Laguna 38200, Tenerife, Spain}

\begin{abstract} We introduce the use of the Fast Multipole Method (FMM) to speed up gravitational lensing ray tracing calculations. The method allows very fast calculation of ray deflections when a large number of deflectors, 
  $N_*$, is involved, while keeping rigorous control on the errors. 
  In particular, we apply this method, in combination with the Inverse Polygon Mapping technique (IPM), to quasar microlensing to generate 
  microlensing magnification maps with very high workloads (high magnification, large size and/or high resolution) that require a very large number of deflectors. Using, FMM-IPM, the computation time  can be reduced by a factor
  $\sim 10^5$ with respect to standard Inverse Ray Shooting, making the use of this algorithm on a personal computer comparable to the use of standard IRS on GPUs.  We also provide a flexible web interface for easy calculation of microlensing magnification maps using FMM-IPM\footnote{\url{https://gloton.ugr.es/microlensing/}}. We exemplify the power of this new method by applying it to some challenging interesting astrophysical scenarios, including clustered primordial black holes, or extremely magnified stars close to the giant arcs of galaxy clusters.
    We also show the performance/use of FMM to calculate ray deflection for a halo resulting from cosmological simulations composed by a large number ($N\gtrsim 10^7$) of elements.
\end{abstract}


\section{Introduction}
\label{intro}

Gravitational lensing constitutes at present a standard tool in astrophysical/cosmological research which allows to study several aspects of the universe (e.g. dark matter content/structure in galaxies/clusters, the abundance of MACHOs, the structure of active galactic nuclei, dark energy, etc..), difficult to probe by other means  (e.g. Mediavilla et al., 2016). In order to compare the theoretical predictions of different models with real observations, computer simulations are very often required. 

One of the most common tasks in computational gravitational lensing is ray-tracing, consisting in the mapping of points from the lens to the source plane through the lens equation. On the other hand, it is also frequent that the mass distribution in the lens that determines this mapping is composed of many particles or not known through an analytic expression but described numerically.  Moreover, for some problems, ray tracing may even need to consider the thickness of the lens by using, for instance, several lens planes with mass distributions calculated numerically from cosmological simulations (e.g. Hilbert et al. 2009, Petkova et al., 2014), so the procedure must be repeated for each lens plane, making the computational load heavier. Having algorithms to perform this mapping through the lens equation in a fast and accurate way is, therefore, very important to be able to effectively address many problems in this field (e.g. Plazas, 2020). 

For simplicity, 
we start considering here the most usual case of the so called {\sl thin lens approximation} in which the thickness of the deflecting mass is negligible compared to its distances to source and observer. In this case, the problem can be treated as a two dimensional problem\footnote{There is no loss of generality, as this treatment can be extended to include a more general case of 3D lenses by considering it as a collection of multiple thin lens planes (e.g. Wambsganss, 1999).}. The lens equation that describes this mapping for a single (thin) lens system composed of many ($N_*$) point lenses of mass $m_i$
is\footnote{This equation can also be used for a continuous mass distribution described numerically on a grid, and can be adapted for a more general case in which particles represent groups of particles/pseudo-particles with a different gravitational effect. Moreover, the possible effect of a smooth background gravitational field characterized by its local convergence $\kappa_s$ and shear $\gamma$, which are either constant of smoothly varying in the region of interest, can also be trivially included into this lens equation.}:
\begin{equation}
  {\bf y}=
  {\bf x}-\sum_{i=1}^{N_*}m_{i}{\bf \frac{(x-x_i)}{|x-x_i|^2}}
\label{eqmicro}
\end{equation}
where $\mathbf{x}$ and $\mathbf{y}$ are the normalized angular coordinates in the image and source plane respectively (i.e. in units of the Einstein radius of a reference mass). If we need to calculate the mapping for $N_x$ points in the image plane, the sum on the right hand side of the equation implies $O(N_*\times N_x)$ operations. For many applications in gravitational lensing, both $N_x$ and $N_*$ can be very high (e.g. extragalactic microlensing maps of high magnification systems, or lensing calculations from cosmological simulations which may even include multi-plane lensing), making the computation of this sum very time consuming. For example, the calculation of a typical microlensing magnification map of $2000\times 2000$ pixels may require calculating the above sum for $N_x\sim 10^{10}$ rays
and, for a map with a size of $L_y\sim 200$ Einstein radii and an average magnification of $\mu \sim 20$, we may have $N_*\sim 10^6$ deflectors, resulting in a huge computational effort.

This type of computational problem is indeed very common in physics.
In the field of astrophysics, for example, a similar sum appears in stellar dynamic N-body simulations to account for the mutual gravitational interactions. The most common approach to manage this heavy computational problem is to provide an approximate estimate of the sum (accurate enough for the physical problem at hand) instead of the exact calculation. The approximation is usually addressed by a {\sl divide and conquer} philosophy. The sum is split into pieces, some of which can be calculated in an approximate but much faster way. For this particular problem, an efficient way to achieve this goal is a spatial division of the computational domain, followed by a joint approximate treatment of the contribution of far away lenses. As the gravitational field (i.e. potential, deflection angle, etc) varies slowly with the impact parameter for long distances, this grouping of far lenses is very efficient. This fact was already noticed and used in the early works of extragalactic gravitational microlensing (e.g. Schneider \& Weiss, 1987). They introduced this division between near and far lenses to ease the calculation of this sum by grouping the latter into large groups that can be treated in an approximate way (by considering only the first terms of the multipolar expansion). Meanwhile, nearby lenses were treated directly. 

A refined version of this approach had already been introduced earlier in the field of stellar dynamics, as it is the basis of the famous Barnes \& Hut (1986) hierarchical treecode algorithm (hereafter B-H), in which space is hierarchically divided into progressively finer regions. This is the strategy behind the well known gravitational microlensing code of Wambsganss (1999), who used a hierarchical treecode with multipolar expansion for the deflection of the far lenses (and also introduced a few additional clever interpolation improvements to speed up calculations even more). A similar approach is also used by Metcalf \& Petkova (2014) in their state of the art gravitational lensing software GLAMER. These hierarchical treecode algorithms are highly optimized, reducing the number of calculations for the above sum to complexity $O(N_x \times \log N_*)$, which is a huge improvement for large values of $N_*$\footnote{This is the same degree of optimization achieved by Particle-Mesh algorithms (Hockney \& Eastwood, 1989) based on the Fast Fourier Transform (FFT), which is also commonly used in this context, albeit with its own specific problems (cf. Metcalf \& Petkova, 2014).}. Moreover, treecodes are very versatile (e.g. can be easily applied to adaptive grids and/or paralellized). The accuracy of the approximate part of the calculation is controlled by a parameter $\delta \sim 1$ (usually named {\sl opening angle} or {\sl accuracy parameter}) which determines which objects can be treated in a joint manner and which need to be included individually. Smaller values of $\delta$ provide a better accuracy in the approximation, at the price of slower execution times. The compromise in choosing the right value of $\delta$ is a bit of an art, requiring some trial and error.  The highest term in the multipolar expansions of particle groups, $p$, could also be used to control the precision of the calculations, although most often it is kept fixed to a value {\sl sufficiently high} for the problem at hand. One major drawback of these algorithms is, indeed, the lack of a rigorous a priori control on the error.

There exists, nevertheless, an algorithm that can do this job in an even more efficient way, and with the important property of keeping strict control on the error at all times. This algorithm, introduced not much later than the B-H algorithm, is the Fast Multipole Method (FMM) by Greengard \& Rokhlin (1987), which can reduce the computational effort of the present problem down to complexity $O(N_x + N_*)$  (although the involved constants may render this formal limit difficult to notice in many practical applications)\footnote{The algorithm is so powerful that it has been included among the top 10 algorithms of the twentieth century (cf. Board \& Schulten, 2000), together with codes as famous the Metropolis algorithm for MonteCarlo simulations or the Fast Fourier Transform.}. The FMM has already been used extensively in astrophysics, including state of the art cosmological simulations with trillions of particles (e.g. PKDGRAV3 by Potter, Stadel \& Teyssier, 2017), but surprisingly enough, and despite the fact that the original Greengard \& Rokhlin (1987) paper describes exactly the same two dimensional problem to be addressed in gravitational lensing described by Equation  \ref{eqmicro} (albeit in the electrostatic language), the FMM has (to our knowledge) never been applied in this field. The present work intends to fill this gap, by introducing this powerful algorithm into the toolbox of computational techniques for the gravitational lensing community. As mentioned above, the two scenarios which can benefit more from this algorithm are extragalactic microlensing and lensing calculations using numerical cosmological simulations. We illustrate the use of the FMM algorithm here by focusing mainly on the first of these applications, although we also explore the performance of FMM in the case of a galaxy lens formed by a very large number of elements coming from a cosmological simulation.

The paper is organized as follows. Section \ref{FMM-GL} describes the FMM method and its applicability to gravitational lensing. A specific implementation for the fast calculation of microlensing magnification maps is included in Section \ref{FMM-IPM}.  Regarding cosmological applications, an example of ray tracing considering a numerical lens from cosmological simulations is briefly addressed in Section \ref{cosmo}. A few applications of the code to specific problems of particular scientific interest of high computational demand, difficult to address with previous algorithms/techniques are shown in Section \ref{examples}.  Benchmarks of the new algorithm and comparisons with some other preexisting microlensing codes are presented in Section \ref{Benchmarks}. Section \ref{conclusions} summarizes the final conclusions.

\section{The Fast Multipole Method for gravitational lensing applications}
\label{FMM-GL}

\subsection{ Ray deflection using the FMM}

The FMM provides a very fast tool to map a large number of points from the image plane to the source plane with the required accuracy for cases in which the lens is described by a large number mass elements, $N_*$.  A brief summary of the method can be found in Appendix \ref{AppA}. The advantages of this method are:

\begin{enumerate}
\item Very fast calculation with optimized complexity scaling as $O((N_x+N_*)\times \log_2(1/\epsilon))$, with $N_x$ being the number of rays, $N_*$ the number of deflectors, and $\epsilon$ the required accuracy.  
\item Rigorous control on the errors. Potential and deflection angles can be calculated to the desired degree of accuracy by using the necessary number of terms in the multipolar expansions, $p$.
\item Target points on which the deflections are calculated do not need to be located on a grid, and are independent of the locations of deflectors. This makes the method very general and well suited for problems needing an adaptive grid and/or special geometries.
\item As target points are independent of the location of deflectors, memory use can be easily managed by splitting the set of target points in the lens plane to be mapped back into smaller chunks to suit the memory needs (e.g. see \ref{FMM-IPM} below).
\end{enumerate}  

The FMM is indeed a very versatile algorithm for which parallel versions suitable for GPUs (e.g. Cruz et al., 2011) and kernel independent versions (e.g. Wang et al., 2001) have been released. Therefore, any ray tracing calculation through a lens which is described numerically at many points can, in principle, obtain a great benefit from the application of the FMM to the deflection calculations. These most obvious examples are that of gravitational lensing computations using cosmological numerical simulations (e.g. GLAMER by Metcalf \& Petkova) and the calculation of extragalactic microlensing magnification maps (e.g. Wambsganss, 2006).  In the next sections we introduce the application of the FMM to both cases, starting from the latter one.

\subsection{High workload microlensing magnification maps}

The calculation of extragalactic gravitational microlensing magnification maps (e.g. Wambsganss, 2006), which may involve a large number of deflectors, and which needs to map a huge number of rays, is probably one of the most computationally demanding examples. Gravitational microlensing of extragalactic sources allows to obtain information on several aspects of lenses and sources difficult to obtain by different means (e.g. the abundance and mass of any kind of compact objects in the lens; the quasar structure at several scales, from the large torus (e.g. Popovic et al., 2020) and the Broad Line Region (e.g. Fian et al., 2021 and references therein), down to the innermost regions of the accretion disc (e.g. Morgan et al., 2010; Jim\'enez-Vicente et al., 2012; Mediavilla et al., 2015)).
These studies are usually based on the estimate of the statistical likelihood of the observed magnification of the source for different values of the parameters of interest (abundance and mass of the microlenses, size and temperature profile of the source, etc.). The likelihood is evaluated from microlensing simulations (typically from microlensing magnification maps (e.g. Schneider \& Weiss, 1987)) which can be a very heavy computational task.

 We can define the workload of a magnification map as a quantity that measures the computational effort needed to calculate the map. This workload depends on the size map on the source plane, its resolution (i.e. in the number of pixels of the map), and on the number of deflectors (as the deflection of each ray is the sum of the deflections produced by each of the deflectors in the lens plane $N_*$). Due to the distortion introduced by the lens mapping, the area to be considered in the image plane (and consequently the number of rays to be shot) is also proportional to the average/macro magnification $\mu$, such that $N_{rays}\propto \mu\, N_{pix}$. Ideally, a magnification map should have enough spatial resolution to sample the source, and a size large enough as to include a large number of microlenses\footnote{If a mixture of microlenses with different masses is considered, in order to have a statistically significant number of the most massive ones, a very large amount of the less massive ones might be necessary.} to avoid sample variance\footnote{In principle, sample variance can be mitigated averaging the histograms of several maps, but cooperative effects at 2 or more bodies, particularly the most massive ones, should be present in each realization.}. The workload of a magnification map can therefore be defined as $\Omega=\mu\times N_{pix}\times N_*$, where $N_{pix}$ is the number of pixels in the map. If the map is calculated in a time $t_{ex}$, the performance of such calculation can therefore be measured as $\Omega/ t_{ex}$.
A typical magnification map (e.g. Vernardos \& Fluke, 2013), with magnification $\mu=10$, number of pixels $N_{pix}=4096\times 4096$ and number of deflectors $N_*=50000$, results in a map workload of $\Omega=1.68 \times 10^{13}$ (with an average execution time of $t_{ex}= 7200s$, it produces a performance of $\Omega/t_{ex}=2.3\times 10^9 s^{-1}$).

There is a number of astrophysical scenarios of high interest that require magnification maps of very high workloads and, consequently, very long computing times, even with powerful hardware. For instance, we face this situation in the case of (i) highly magnified images by factors that can range from several tens (multiple imaged quasars, e.g. Jim\'enez-Vicente et al., 2012, 2015) to several thousands (stars close to giant arcs in galaxy clusters, e.g. Welch et al., 2022), which imply both, huge regions to be mapped in the image plane and, consequently, very large numbers of deflectors, (ii) mixed populations combining deflectors of very different masses (e.g., strongly bimodal distributions including massive black holes or mass functions including a large number of substellar and planetary mass objects, e.g. Esteban-Guti\'errez et al., 2022a, 2022b), which force a very large number of small mass particles to have an statistically significant number of the largest mass ones, (iii) clustered/non-uniform lens distributions (e.g., clustered Primordial Black Holes (hereafter PBHs)), which imprint on the maps features associated to the individual particles and to the clusters that act like large mass pseudo-particles and, hence, force also a very large number of individual particles to preserve the statistics of the clusters (other kind of bimodality), (iv) simultaneous study of regions of very different sizes (e.g., from the tiny x-ray emitting region and the UV continuum of the accretion disk to the Broad Line Region (BLR)  and the dusty torus in Active Galactic Nuclei (AGN)), which imply a large combined effort in resolution, size and number of deflectors (markedly pronounced in the case of highly magnified quasars) and (v) time variable magnification maps taking into account the motion of deflectors.  In the next section we review the most frequently used techniques and strategies followed to compute high workload magnification maps.

\subsection{Methods and algorithms to calculate microlensing magnification maps}

 The classical, {\sl brute force} method to compute magnification maps is Inverse Ray Shooting (hereafter IRS), which transports backwards a set of points covering the image plane (rays) to the source plane through the lens equation (cf. Kayser, Refsdal \& Stabell, 1986; Schneider \& Weiss, 1987).
The magnification of each source-plane pixel is made proportional to the number of rays that hit the pixel. Magnification estimates from IRS have an inherent noise, which is Poissonian for random rays or somewhat smaller for rays on a grid (Kayser, Refsdal \ Stabell, 1986, Mediavilla et al., 2011). In order to reduce that noise, a high number of rays per element of resolution in the absence of lensing (typically $n_0 \sim 500$) needs to be shot. 

Alternatively, in the Inverse Polygon Mapping (IPM) technique (Mediavilla et al. 2006, Mediavilla et al. 2011),the cells/tiles of a regular tesellation of the image plane are transported backwards to the source plane using inverse lens mapping. The magnification for a pixel in the source plane of area $S^*$ is proportional to the ratio of the areas of all its images and the area of the pixel (i.e. $\mu=\sum{S_i}/S^*$). The IPM directly uses this definition and, for each transported cell in the image plane, Green's theorem is used to exactly apportion its area among the source-plane pixels covered by the transformed cell. All cells (or fractions) in the image plane covering the source plane pixel therefore contribute to the numerator in the sum above. Critical cells (containing a critical curve) can be treated with different degrees of precision (cf. Mediavilla et al. (2011)), although a naive treatment (as if they were non-critical) produces the correct result for most applications. A great advantage of this procedure is that, in the low magnification regions, which for standard IRS would need a large number of rays to be shot to reduce the noise, the IPM provides the correct magnification for very few rays, as the inverse mapping is a very smooth mapping in these regions.
This allows this method to work with $n_0 \lesssim 1$ while producing maps with very low noise, comparable to IRS maps with $n_o\sim 500$. As the computational time for calculating a microlensing magnification map is proportional to the number of rays shot per unlensed pixel $n_0$, a factor of a few hundred can therefore be saved by using IPM techniques over plain IRS.

 As commented above, another improvement, the separation of far and near gravitational fields to speed up the calculations, was introduced quite early (cf. Schneider \& Weiss, 1987) and subsequently refined using the Barnes \& Hut (1986) hierarchical treecode (hereafter B-H) concept and some further improvements by Wambsganss (1999), resulting in a very strong reduction of the computational times for large number of deflectors $N_*$. An alternate path to address this issue has been using Particle-Mesh methods in combination with FFT techniques to solve Poisson equation on a grid (e.g. Kochanek, 2004) to calculate the far field contribution to the potential and the corresponding deflection field.

Irrespective of the improvements in the algorithm, using faster hardware like supercomputers (e.g. Garsden \& Lewis 2010) or Graphic Processing Units (GPUs, e.g. Thompson et al., 2010) that take advantage of the highly parallelizable nature of the microlensing problem, has also been an alternative approach to produce magnification maps in reasonably short execution times, allowing the mass production of large datasets of maps (e.g. Gerlumph database by Vernardos et al., 2014). These databases are a huge asset for addressing some problems, but many possible interesting models/scenarios are not included in these simulations (i.e. different mass functions including strongly bimodal mixed populations for the deflectors, non-uniform spatial distributions, extreme magnifications and or resolutions, etc.).

A logical step to face  the computation of high workload magnification maps is to combine the performance benefits of the IPM approach with an efficient algorithm to rapidly evaluate the ray deflections produced by systems with a high number of deflectors. Despite this solution has long been suggested (e.g. Mediavilla et al., 2011; Jim\'enez-Vicente, 2016), it has only been carried out recently. Shalyapin et al. (2021) have been the first to follow this path by implementing a Poisson Solver for calculating the ray deflections in an IPM based code named Poisson and Inverse Polygon (PIP). The resulting software is very efficient in generating microlensing magnification maps. They have also provided a web interface for their code that can produce magnification maps for reasonably low magnifications ($\mu < 50$) at moderate resolutions in very short times, even if a significant number of deflectors is required. 
This is without doubt a very significant step forward, but although using a Poisson solver to calculate the deflection of rays on a 2D grid is a computationally efficient approach (as it makes use of the very fast FFT algorithm), particularly for the case of many deflectors, it carries some inherent inconvenients which may  make this procedure unsuitable for some very demanding applications like the ones mentioned above.
In particular, the solution to the Poisson equation has to be calculated simultaneously for a whole regular grid of rays. This means that the deflections for all the rays in the image plane need to be computed at once. This is a rigidity that can easily overflow the memory of many computers when the number of rays to be considered is very high, either because the image plane is very large (e.g. due to high magnifications) and/or because the need of high spatial resolution. For example, Shalyapin et al. (2021) consider a case of $\mu=2.5$ and $N_{pix}=2000\times 2000$, resulting in an image plane grid made up of $7500\times3000=2.25\times10^7$ rays. On the other hand, considering a more demanding case $\mu=100$ and a map of $N_{pix}=4000\times 4000$ pixels, would require to solve the Poisson equation on a grid of $600000\times 6000=3.6\times 10^9$ rays, which would require $\sim 160$ times more memory. Moreover, the method may not be easily used if adaptive grids or very high resolutions are needed.

The current implementation of the PIP uses a rectangular distribution of the deflectors in the lens plane. This approach may show only a slight impact in the borders of the maps when moderate magnification is considered, but can produce strong field distortions (pincushion/barrel) in the magnification maps for high magnification cases, particularly for large values of $\kappa_*$ (e.g. a map of 50 $E_R$ with $\kappa=\kappa_*=1.2$, $\gamma=0.4$ shows obvious distortions.). According to the  2D shell theorem, the deflectors should be distributed in a circular region circumscribing the rectangular ray shooting region to prevent these field distortions. This can be corrected either analytically (cf. Zheng et al. 2022) or by explicitly including the circular distribution of deflectors, at a price of extra computational time (which can be significant for extreme magnifications which demand very large radii for the circle). The latter option will largely benefit from the implementation of the FMM.

Another inherent inconvenient of this approach is that in these Particle Mesh methods, lenses should be {\sl re-gridded} to the image plane grid of rays. Shalyapin et al. (2021) have chosen to use a standard {\sl cloud-in-cell} method for this step (Hockney \& Eastwood, 1989), and they acknowledge that it works reasonably well, but they also mention that alternative methods as the {\sl triangular shape cloud} produces smoother mass distributions that generate unrealistically smooth magnification maps. While the {\sl cloud-in-cell} method can produce good results for many cases, it is not completely clear what is the effect of this approximation in general (e.g. strongly bimodal mass spectrum of the microlenses and/or extremely high/low resolutions). There is no much a priori control on the accuracy of this approximation, and on its dependence on resolution.

 In the next section, we present our proposal to combine IPM with FMM to avoid these drawbacks.

\subsection{A fast FMM-IPM  algorithm to calculate microlensing magnification maps}
\label{FMM-IPM}

  Our algorithm uses the FMM for the calculation of the ray deflections with the desired precision, and the IPM to optimize the number of rays to be shot, to produce high quality maps with very low noise (comparable to IRS maps with $n_o\sim 500$ rays per unlensed pixel).

Although the IPM method can be set to work at a higher order by making a careful treatment of critical cells for very high precision maps near caustics (cf. Mediavilla et al., 2011), from the practical point of view, this is not necessary for most applications. We therefore will use here the original first order (being IRS the zeroth order approximation) IPM code as described by Mediavilla et al. (2006).

In order to keep memory usage under control, and using the idea described above, a parameter \texttt{nmaxarg} is defined to control the maximum number of rays to be shot at once. Moreover, and in order to keep maximum versatility in the code that calculates the magnification map, we have kept independent from it the previous task of setting up the distribution of lenses in the lens plane. The structure of the code is therefore very simple:

\begin{enumerate}
\item Read the input parameters: Macro model parameters ($\kappa$, $\gamma$), mass in form of compact objects ($\kappa_*$), map size in Einstein radii ($y_l$) and in pixels ($N_y$), distribution of deflectors in the mass plane ${\bf x_i}$ and their masses ($m_i$), number of rays per unlensed pixel ($n_o$), and accuracy of the FMM ($\epsilon$).
\item Partition the image plane into $n_c$ chunks ($c_i$) each containing at most \texttt{nmaxarg} rays. $n_c \sim N_{rays}/$\texttt{nmaxarg}.
\item For each chunk $c_i$ ($i=1, \ldots, n_c)$: calculate the deflected rays ${y_k}$ for all rays ${x_k}\in c_i$ using FMM. Include the effect of the background smooth gravitational field.
\item With the mapped polygons formed by deflected rays ${y_k}$, use IPM to apportion the corresponding areas to pixels in the source plane magnification map.
\item Write the magnification map to a file.  
\end{enumerate}

For the calculations carried out in this paper, we have set the parameter \texttt{nmaxarg}=$1.6\times 10^7$. This keeps memory usage under $\sim$ 2 Gigabytes for most cases\footnote{If a very large number of lenses $N_*\gtrsim 10^7$ is used, memory use may be dominated by the tree structure.}, making it suitable for standard personal computers or cluster nodes with modest memory capabilities. 

For the FMM part of the code, we have made use of available libraries. As the original IPM code was written in Fortran, we have made use of two different FMM-2D Fortran libraries with very similar results\footnote{We have used the original FMM2DLIB library by Gimbutas \& Greengard from \url{https://cims.nyu.edu/cmcl/fmm2dlib/fmm2dlib.html}, and its updated version FMM2D from the Flatiron Institute in \url{https://github.com/flatironinstitute/fmm2d} with no significant difference in performance.}.

\begin{figure*}[htb]
\includegraphics[width=18cm]{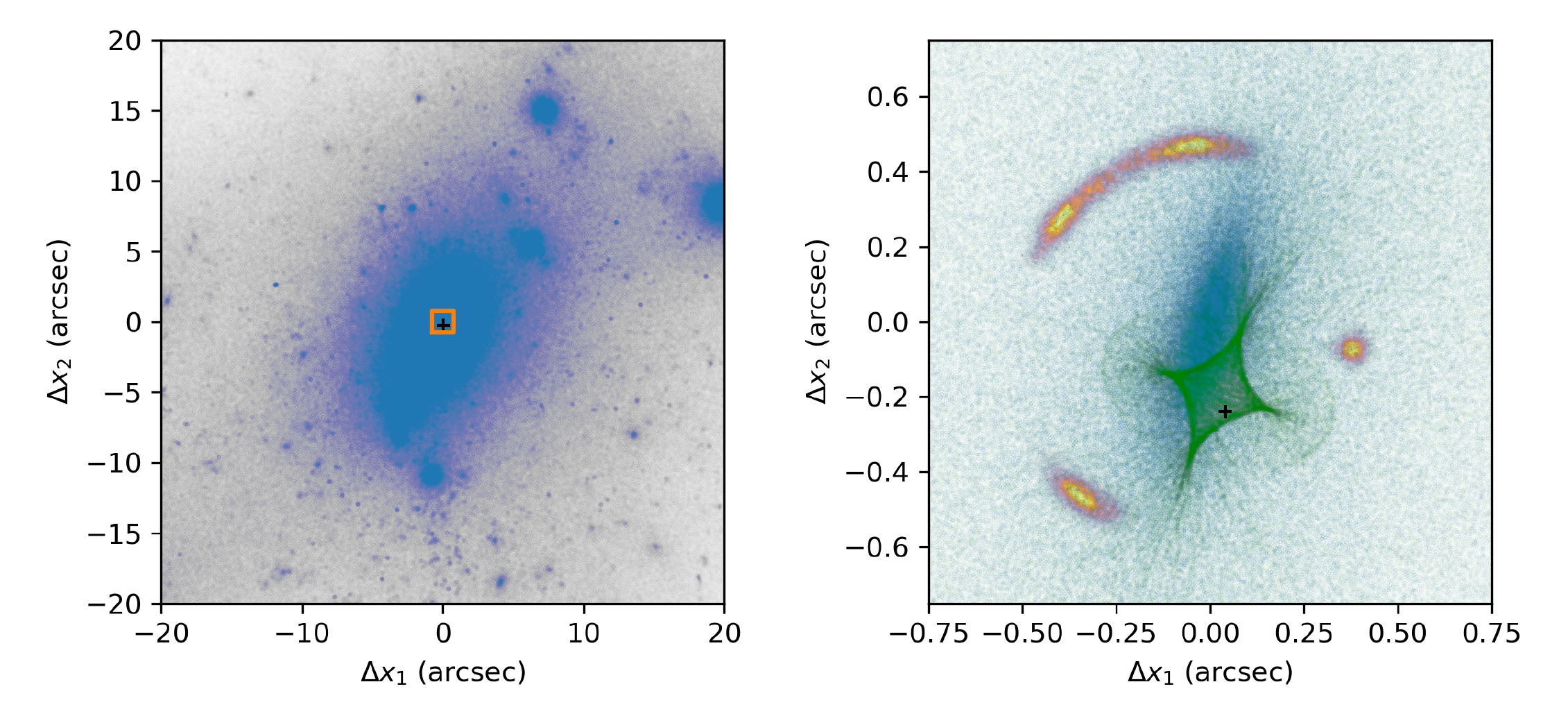}
\caption{Lensing by a galaxy halo from the EAGLE simulation with $15.75\times 10^6$ particles. Left panel shows the lens particle distribution in the innermost 40 arcsec. A zoom of the region indicated with the orange square is shown on the right panel. The location of a background source is indicated as a black cross. Right panel shows the innermost 1.5 arcsec including lens particles (in blue), source location (black cross), magnification distribution in the source plane (in green) and the multiple images of the source.\label{halo}}
\end{figure*}

Although the precision in the deflection calculation can be adjusted/optimized for the problem at hand, for the calculations presented in this paper we have made a very conservative choice to keep the error fixed to $\epsilon=5 \times 10^{-5}$, which is indeed a very high accuracy\footnote{Taking into account that the Einstein radius of the reference mass is the natural length unit in the source plane, and therefore, pixels have typical sizes $\gtrsim 0.001 E_R$}, well above what is usually needed for most cases/applications. With this settings, the execution time for the calculation of a magnification map involving a large number of lenses $N_*\gtrsim 10^5$ can be reduced by a factor $\sim 500$ with respect to the standard IPM,  allowing the calculation of heavy magnification maps with a standard personal computer in much shorter execution times (see Section \ref{Benchmarks} below).

At present, and despite the used FMM libraries can benefit from multicore processors via Open Multiple Processing (OMP), we have made no attempt to use multithreading capabilities in our code, so a single core is used in the presented calculations. In the present work, all the simulations have been performed on a standard desktop personal computer with an Intel Xeon W-2123 processor at 3.60GHz.

 Shalyapin et al. (2021) provided a web interface for their PIP algorithm, which allows researchers to easily calculate magnification maps (with some limitations like $\mu<50$). We have followed this excellent initiative, and we have created a similar web interface for our code. In order to prevent the introduction of just a redundant service, we have intended to overcome some of the limitations of the PIP code and interface, and we have tried to make our offer somewhat more flexible for the users (including, for example, the possibility of uploading their own lens plane configuration and larger macro magnifications). This new web service\footnote{\url{https://gloton.ugr.es/microlensing/}} should allow interested researchers to produce magnification maps for a given set of parameters, even for very high workloads.
Researchers interested in using this code for other specific cases not covered by this interface, or for the production of a large number of magnification maps, are very welcome to contact the authors for scientific collaboration.

\subsection{FMM applied to a numerical lens from a Cosmological simulation}
\label{cosmo}

As mentioned before, besides microlensing magnification maps, other type of lensing calculations can also greatly benefit from the use of FMM. In this section, we present a simple example of the use of FMM to a rather general case of ray deflection by a lens galaxy halo. We have taken the galaxy halo from the EAGLE cosmological simulation (Schaye et al. 2015). The galaxy lens has a mass of $M_{200}=10^{13.09} M_\odot$ and a size of $R_{200}=406.32$ kpc, and it is located at a redshift $z=0.5$. It is made up of $15.75\times 10^6$ particles of which $3.76 \times 10^6$ particles are of gas, $10.8 \times 10^6$ particles are dark matter, $1.19 \times 10^6$ are stars, and there are 128 black hole particles.

The left panel of Figure \ref{halo} shows as blue dots the particle distribution (we have made no distinction between the different types of particles for clarity). We use the FMM algorithm to calculate the deflection of a regular grid of $4000 \times 4000$ rays by this lens galaxy in the innermost region (shown in the right panel). The calculation is very fast, taking only $\sim 45$ seconds. 

We have calculated the image of a background source behind this halo using these deflections. The right panel of Figure \ref{halo} shows the inner 1.5 arcsec. The lens particle distribution is again shown as blue dots, ray hits in the source plane (tracing magnification in the source plane) are shown as green dots (clearly showing the tangential and radial caustics), and the location of a background (Gaussian) source is shown as a black cross. The four images of this source are overplotted in this panel.

This is only an example intended to show the potential of FMM in a more general calculation involving a large number of particles outside the gravitational microlensing field (which is the main focus of the present work). We are aware of some specific issues than need to be addressed in this context and which, being beyond the scope of this work, are therefore not dealt with in this example. 

\section{Selected Applications}
\label{examples}

In order to show the capabilities of the microlensing code described above, we present in this section the results from the application of the FMM-IPM code to a few selected cases of current high scientific interest which require a large number of lenses $N_*$ and/or a large number of rays, and consequently with high workloads $\Omega\gtrsim 10^{15}$. These problems would therefore be difficult to address with preexisting codes without high performance/memory hardware, and/or too long execution times. 

\subsection{Quasar structure from the event horizon to the torus: the case of the highly magnified image B of quasar PSJ 0630-1201}

\begin{figure}[htb]
\includegraphics[width=\linewidth]{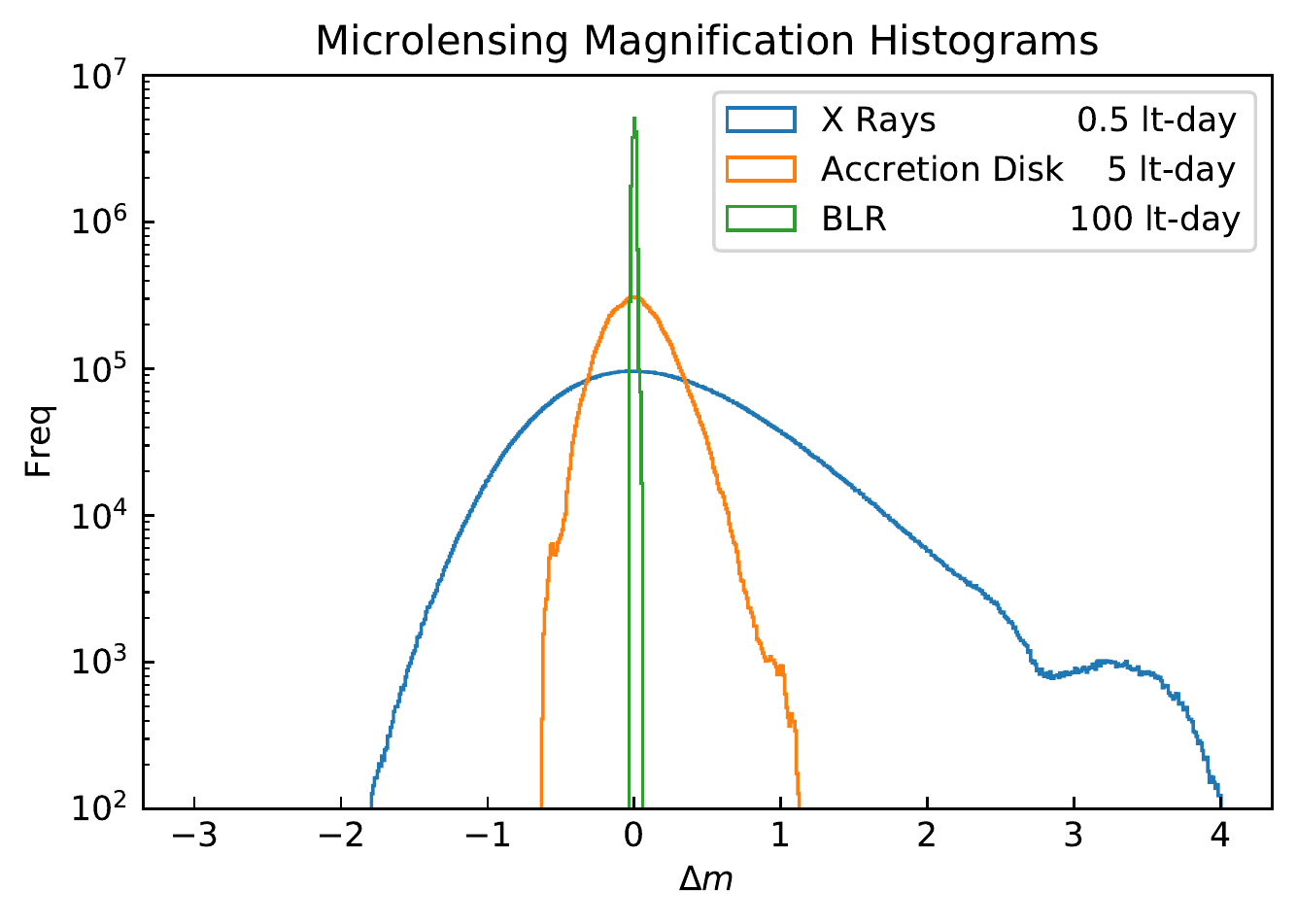}
\caption{Microlensing magnification histograms for image B of lens system PSJ 0630-1201 (Shajib et al. 2019) for different source sizes. Blue line corresponds to the X-Rays ($r_s=1$ lt-day), orange line to the UV continuum ($r_s=5$ lt-days) and green to the BLR ($r_s=100$ lt-days). The Einstein radius (for microlenses of 0.2 $M_\odot$) is $R_E$=8.5 lt-days. \label{blrhist} }
\end{figure}

Microlensing and millilensing affect with different strengths the regions in which AGN and quasars are structured depending on their size: from the smallest scales of the event horizon and the Innermost Stable Circular Orbit (ISCO, cf. Mediavilla et al. 2015), through the intermediate scales of the accretion disc and BLR (e.g., Morgan et al. 2010, Jim\'enez-Vicente et al. 2012, Fian et al.  2021), up to the large scale of the torus of scattered light (cf. Popovic et al. 2020). For single epoch microlensing, as the intrinsic brightness of the source is unknown, a baseline defining zero microlensing is needed to measure the impact of microlensing magnification (e.g., Mediavilla et al. 2009). This reference is usually taken from the emission coming from a region of the quasar which is large enough as to be insensitive to microlensing (typically the BLR).
Then, we need to simulate microlensing observations covering large regions of the source plane (to contain several times the largest considered emitting region) while simultaneously keeping high spatial resolution (to be able to probe the smallest relevant scales), which impose very demanding conditions on the magnification maps. To show this, we present here the case of image B in the high magnified lensed quasar PSJ 0630-1201 (Shajib et al. 2019). We have considered an spatial resolution of 0.475 light-days, enough to sample the ISCO in a $\sim 10^9M_\odot$ black hole, and a size of 1900 light-days, which covers the torus of scattered light in a typical case (Popovic et al. 2020). The map has $\kappa=0.49$ and $\gamma=0.52$, with a magnification $\mu=-97$, and $N_{pix}=4000\times 4000$ pixels, which implies a large number of deflectors (taken to have $0.2M_\odot$) $N_*=1.4\times 10^7$. The workload for this map is $\Omega=2.15 \times 10^{16}$. The execution time with the FMM-IPM code was roughly one hour ($t_{ex}=3986 s$). The high number of lenses involved would have made this map a very heavy task even for GPUs (e.g. see typical execution times and performances for Gerlumph in Figures \ref{etvswl} and \ref{pvswl}). Even for IPM, the map needs to shoot a huge amount of $N_{rays}\sim 3.5 \times 10^9$ rays, which is equivalent to an array of $\sim 600000\times 6000$ pixels (which would imply very high memory requirements for a Poisson solver).

\begin{figure*}[htb]
  \includegraphics[width=18cm]{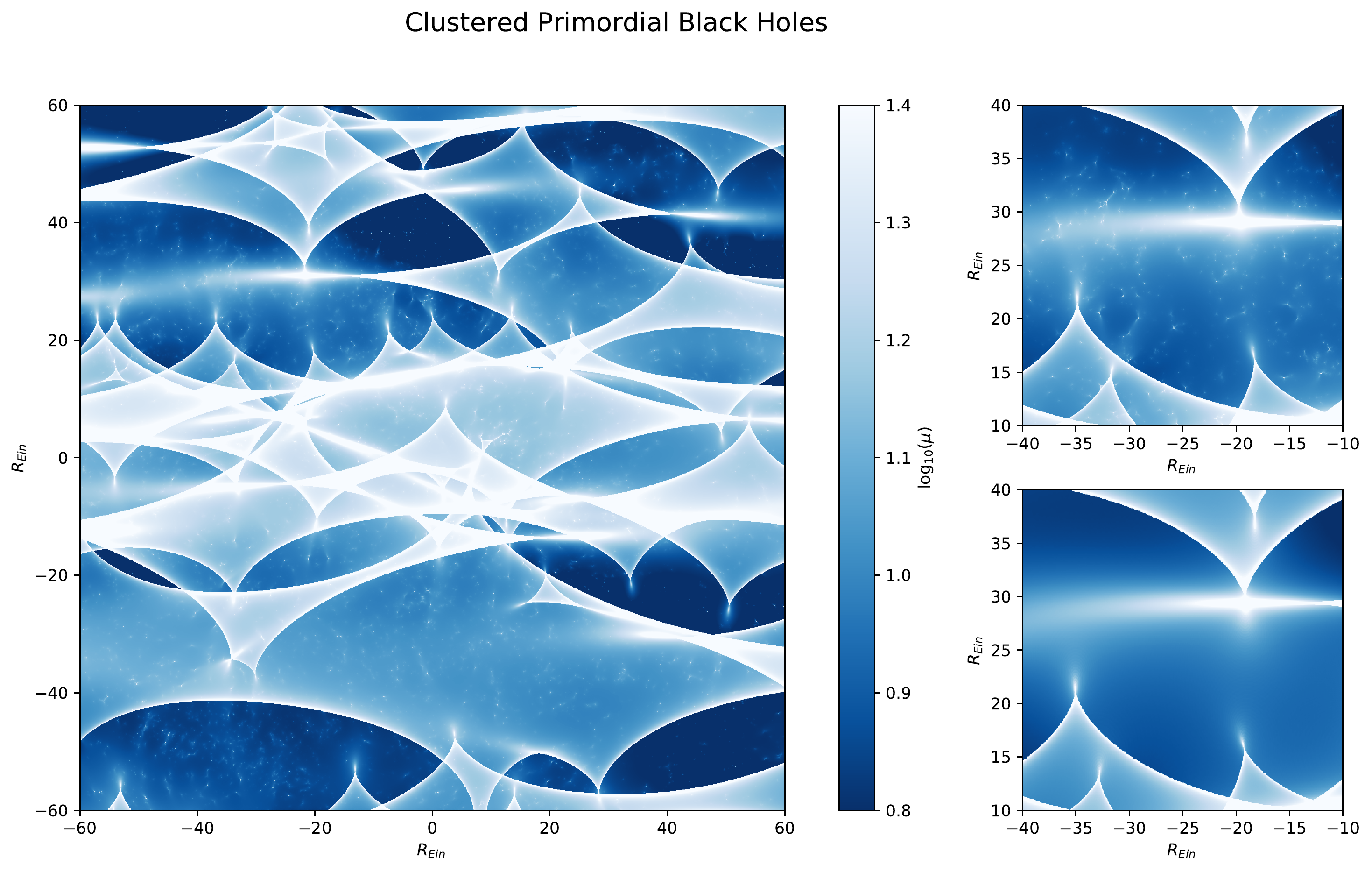}
  \caption{Magnification map for image A1 of system MG0414+0534 produced by 20\% of the projected mass in form of clustered Primordial Black Holes. Insets on the right show a zoom-in comparing of the full simulation including individual BHs  (upper) and considering the clusters as a single object (lower). 
    \label{clusters}}
\end{figure*}

In Figure \ref{blrhist} we present the magnification histograms corresponding to three regions of significantly different sizes, from the innermost regions traced by emission in X-rays to the BLR. It is worth noting the low noise in the histograms (due to the large size of the map). From the histograms, it can be clearly seen that, while the smallest scales (ISCO and accretion disc) are strongly magnified, the BLR and the torus of scattered light are rather insensitive to microlensing and may be used as baseline for no microlensing. These large high resolution maps are necessary to study simultaneously the effect of microlensing on the different sizes involved in the structure of quasars.

\subsection{Inhomogeneous spatial distributions of deflectors: the case of clustered PBHs}

We have so far considered that the deflectors are homogeneously distributed (i.e. that the statistical properties of the deflectors spatial distribution are similar at every point). However, studies about the origin of PBHs (an interesting astrophysical candidate to explain, at least partially, the dark matter) suggest that they form in clusters (e.g. Garc\'{\i}a-Bellido \& Clesse, 2018). In addition, clustering introduces in the problem two spatial scales: the Einstein radius of the BHs and that of the clusters, which behave as pseudo-particles. While the latter are homogeneously distributed in the lens plane, the former are not. The strong inhomogeneity may be a challenge to methods based in the calculation of the gravitational potential from the redistribution of the masses in a regular lattice for two scales are now present.
Figure \ref{clusters} shows the magnification map of image A1 of the system MG0414+0534 where 20\% of the mass surface density is in the form of clustered PBHs of $30 M_\odot$. The simulation contains 1607 clusters with a size of 1pc, each one containing 300 PBHs (for a total of 482237 PBHs). The map is $8000\times 8000$ pixels, for a total workload of $\Omega=4.8\times 10^{14}$. The insets on the right panel show a zoom in of a region of the map for the case including the individual PBHs (upper panel), and also for the case including only the clusters as pseudo-lenses (lower panel). The comparison clearly shows the difference, and the need of including the individual PBHs, which can produce individual caustics that may be present in the observed light curves, but would be absent if they are not included in the simulations. In a forthcoming paper on this issue (Heydenreich et al., private comm.) we have performed simulations of this type which, for large values of the fraction of mass in form of compact objects, require huge number of deflectors up to $8\times 10^7$ which the FMM-IPM code has been able to handle comfortably in much shorter execution times.

\subsection{Extremely magnified stars by galaxy clusters: the case of Earendel}

\begin{figure*}[htb]
  \includegraphics[width=18cm]{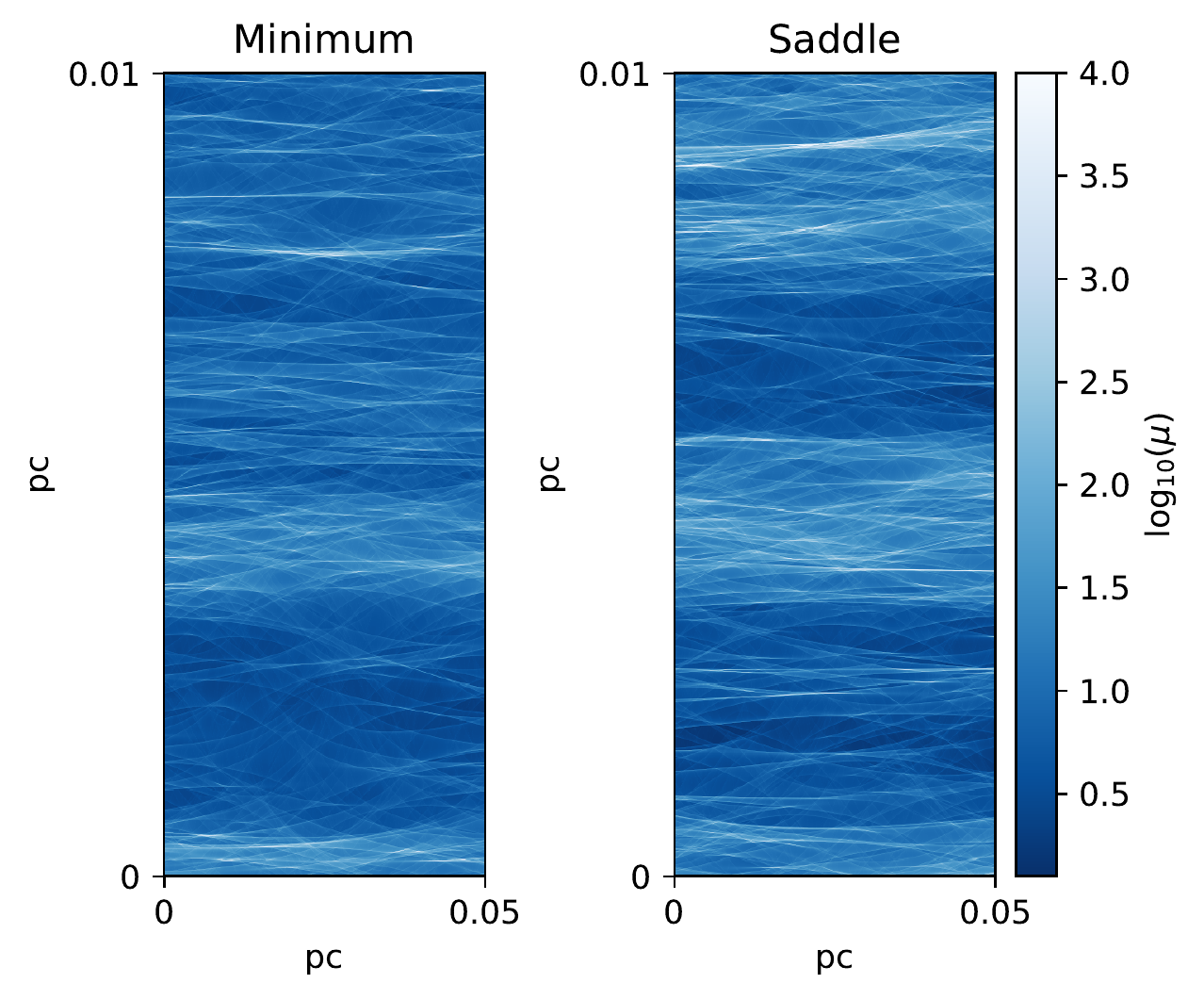}
\caption{Magnification maps for the minimum (left) and saddle (right) images corresponding to Earendel. Vertical axes have been expanded for clarity. \label{Eamaps}}
\end{figure*}

\begin{figure*}[htb]
\includegraphics[width=18cm]{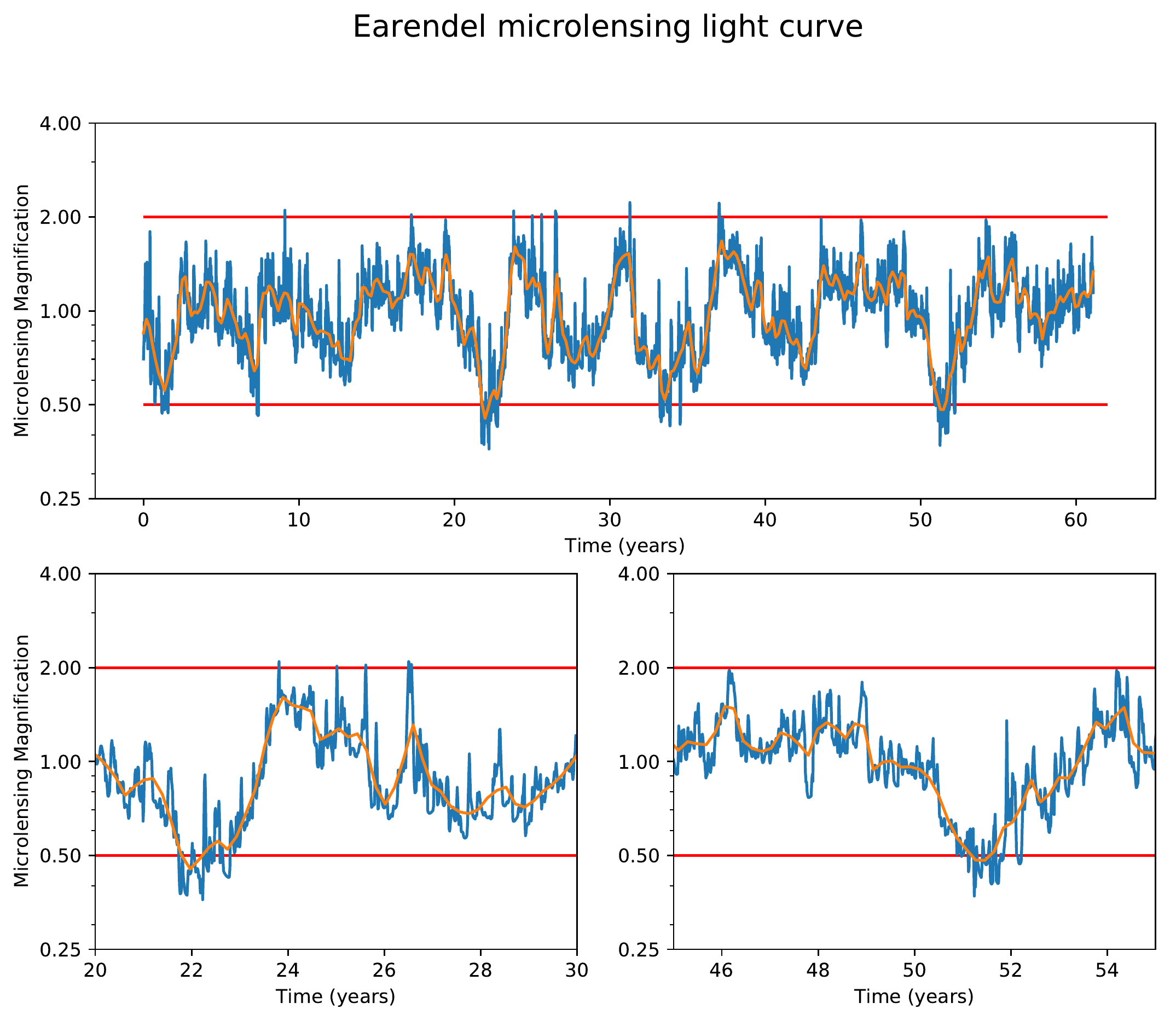}
\caption{Light curve for Earendel for a period of 60 years for sources of $500 R_\odot$ (blue) and $8000 R_\odot$ (orange). Bottom panels show a zoom-in of selected regions. \label{Eacurve}}
\end{figure*}

\begin{figure}[htb]
\includegraphics[width=\linewidth]{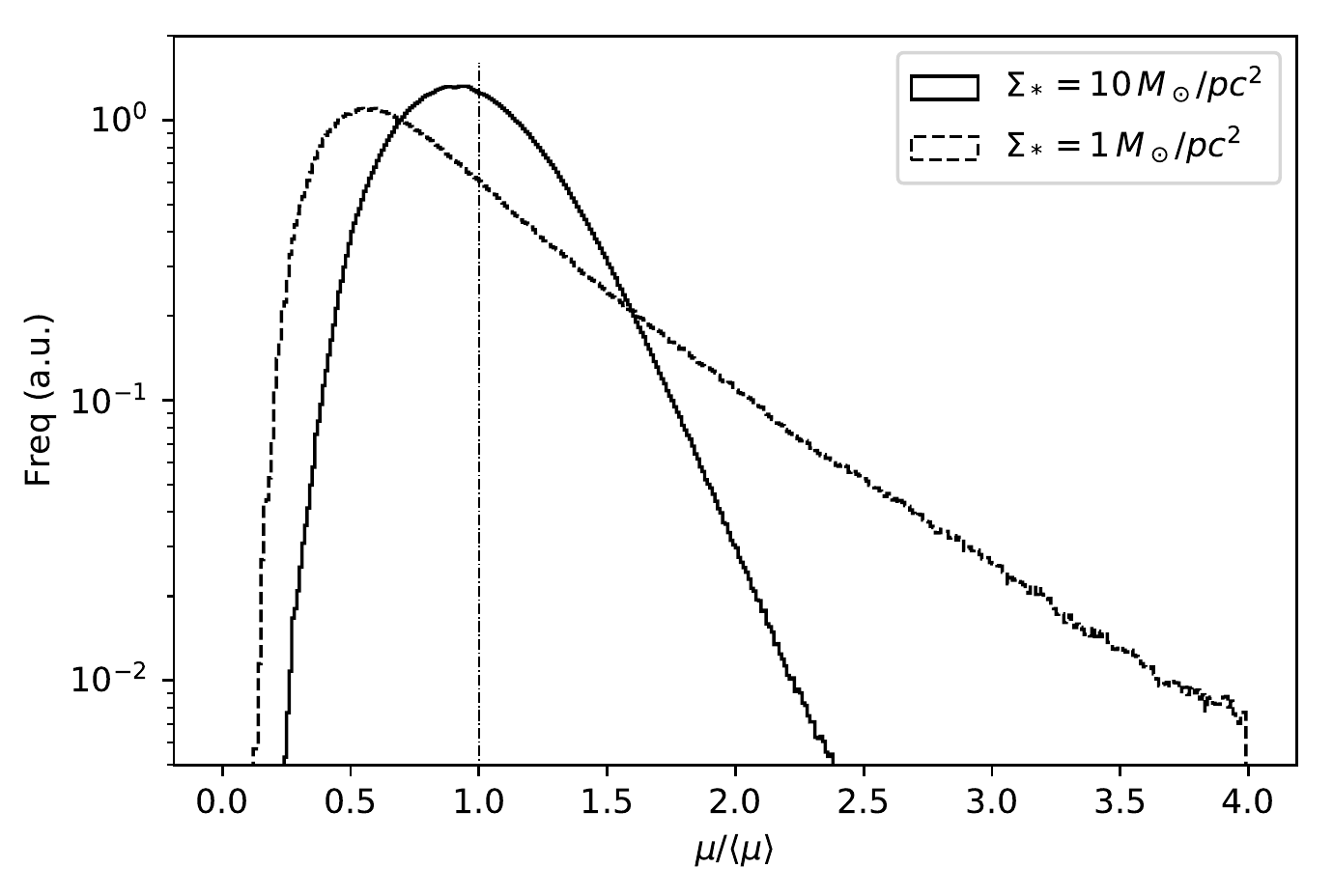}
\caption{Microlensing magnification histograms for Earendel for sources of $500 R_\odot$ for stellar densities of $10\, M_\odot/pc^2$ and $1\, M_\odot/pc^2$. \label{Eahist}}
\end{figure}

Individual stars of far away galaxies can fortuitously align with an intervening cluster of galaxies, experimenting magnifications by factors of thousands (Kelly et al., 2018, Welch et al., 2022). The images of these stars are prone to microlensing by deflectors in the intracluster medium, which can induce microlensing variability. We are going to consider here the case of Earendel, a $z=6.2$ individual star magnified by a cluster at $z=0.566$ discovered by Welch et al. (2022). They interpret the observed unresolved source on top of the arc as the sum of two very close images, each at one side of the arc (two images of the same brightness of positive and negative parity at the same distance of the critical curve according to the local expansion of the potential). The LTM macrolens model considered by Welch et al. (2022) assigns a magnification of 4200 to each of the images with a ratio of 1500 between the magnifications of both  axes. To properly calculate a magnification map of $4\times 4$ Einstein radii (equivalent to 0.05 pc for 1 solar mass deflectors), needed to track the variability of the star along 60 years (at a velocity of 1000 km/s according to Welch et al. 2022) with an angle of 45 degrees respect to the caustic, a very anamorphic rectangular region of $15000 \times 10$ Einstein radii needs to be traced backwards. If $n_0=1$ rays per unlensed pixels are shot, for a resolution of 0.001 Einstein radii per pixel, this corresponds to an array of $15000000\times 10000$ rays, which is unmanageable for a Poisson solver in most computers. Adopting a projected density of 10 $M_\odot/pc^2$ (Welch et al. 2022), we need to consider nearly two millions deflectors (which we have chosen to be of $0.2 M_\odot$) distributed in a huge circular region with a diameter of at least 15000 Einstein radii (roughly 200 pc). 
Finally, a very high spatial resolution of 0.001 Einstein radii (approximately 500 solar radius) to approximately match the source size is also required, preventing an artificial smoothing of the features of the magnification map.  Thus, our maps have $\mu=\pm 4200$, $N_{pix}=4000^2$ pixels, and $N_*=1.8\times 10^6$ lenses, resulting in a map performance, $\Omega=1.2 \times 10^{17}$ which took roughly one day ($t_{ex}\sim 77500 s$) to be computed.

In Figure \ref{Eamaps} we present the magnification maps for both images, which show an structure of clearly resolved stripes (compressed caustics), with the expected larger regions of low magnification in the case of the negative parity (saddle point) image. In Figure \ref{Eacurve} we present the light curve of the star (as the sum of the two images) which shows strong changes in time scales of years. In the same Figure we have also represented a light curve with a lower resolution of {\sl only} $8000 R_\odot$/0.2 light-days, which shows the artificial smoothing induced by the use of an insufficient resolution in the magnification map. Despite a reasonable resemblance with the light curve published by Welch et al. (2022), the light curve in Figure \ref{Eacurve} (even at our lowest resolution) shows more oscillations around the average magnification, with frequent variations of a factor of $\sim$ 3 over periods of roughly two years, which are not present in their curve (which remains stable above mean magnification during most of the 60 year period). It is nevertheless difficult to judge the origin of this discrepancy. At the same time, while they claim their small scale oscillations to be due to shot noise in the ray tracing calculations, variations at the smallest scale in Figure \ref{Eacurve} are real, with frequent variations of magnification of $\sim$ 2 with a time scale of weeks/months (e.g. peaks in the high resolution curve around years 25,26 or 52).

Several algorithms for the calculation of microlensing light curves, particularly well suited for these high magnification cases, have appeared recently in the literature (Venumadhav et al., 2017; Diego et al., 2018; Meena et al., 2022; Diego, 2022). For most applications, a statistical analysis containing many (possibly millions) random light curves at different orientations are usually needed to compare with observations, and therefore the advantage of single light curve calculation over full magnification maps (which can provide many light curves at once) is arguable. 
In any case, all of these new procedures could easily obtain an important performance benefit by using the FMM in their deflection calculations. Conversely, our code could easily be adapted to produce a much smaller map at high resolution of a rectangular region of the source plane containing the track producing the desired light curve. 
In Figure \ref{Eahist} we show the microlensing magnification histograms for Earendel (sum of the minimum and saddle images) for sources of $500 R_\odot$ for stellar densities of $10\, M_\odot/pc^2$ and  $1\, M_\odot/pc^2$ calculated using $8\times 10^8$ events. We see that extreme magnifications are much more common in the latter case. For example, in the former case, only 4\% of the cases have magnifications outside the range 0.5 - 2 with respect to the average, but this frequency increases to 28\% for the latter case. This histogram is easily calculated with magnification maps, but would need to generate many thousands of different light curves with different random orientations to produce a similar result.

\section{Benchmarks and Comparisons}
\label{Benchmarks}

In this section we intend to show the performance of the introduced microlensing FMM-IPM code for different cases and its scaling with different parameters (e.g. number of lenses and map size). We find also interesting to show some brief comparisons with other existent codes. As we are interested in the comparison of the newly introduced FMM algorithm, we have included here only codes also using IPM for the magnification maps (as IPM already introduces a large performance gain with respect to IRS which we do not want to influence the comparison). All simulations have been performed with a very conservative  accuracy of $\epsilon=5 \times 10^{-5}$ in the calculation of ray deflections.

\begin{figure*}[htb]
\includegraphics[width=18cm]{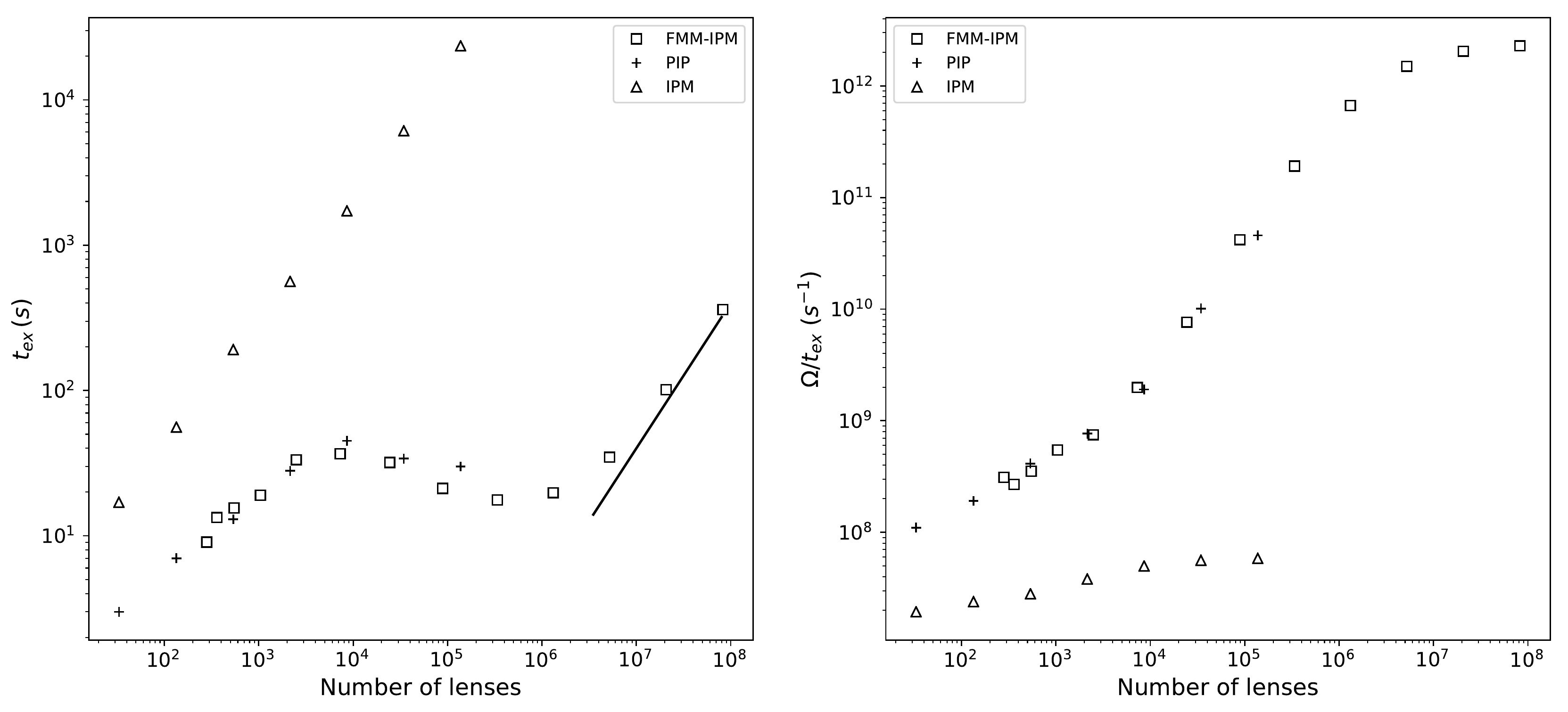}
\caption{Execution time (left) and Performance (right) vs number of deflectors for the FMM-IPM compared to other algorithms. Maps were made of $2000\times2000$ pixels, for $\kappa=\gamma=0.3$, with $\mu=2.5$. The straight line indicates linear scaling with the number of deflectors. \label{pwns}}
\end{figure*}

\begin{figure}[htb]
\includegraphics[width=\linewidth]{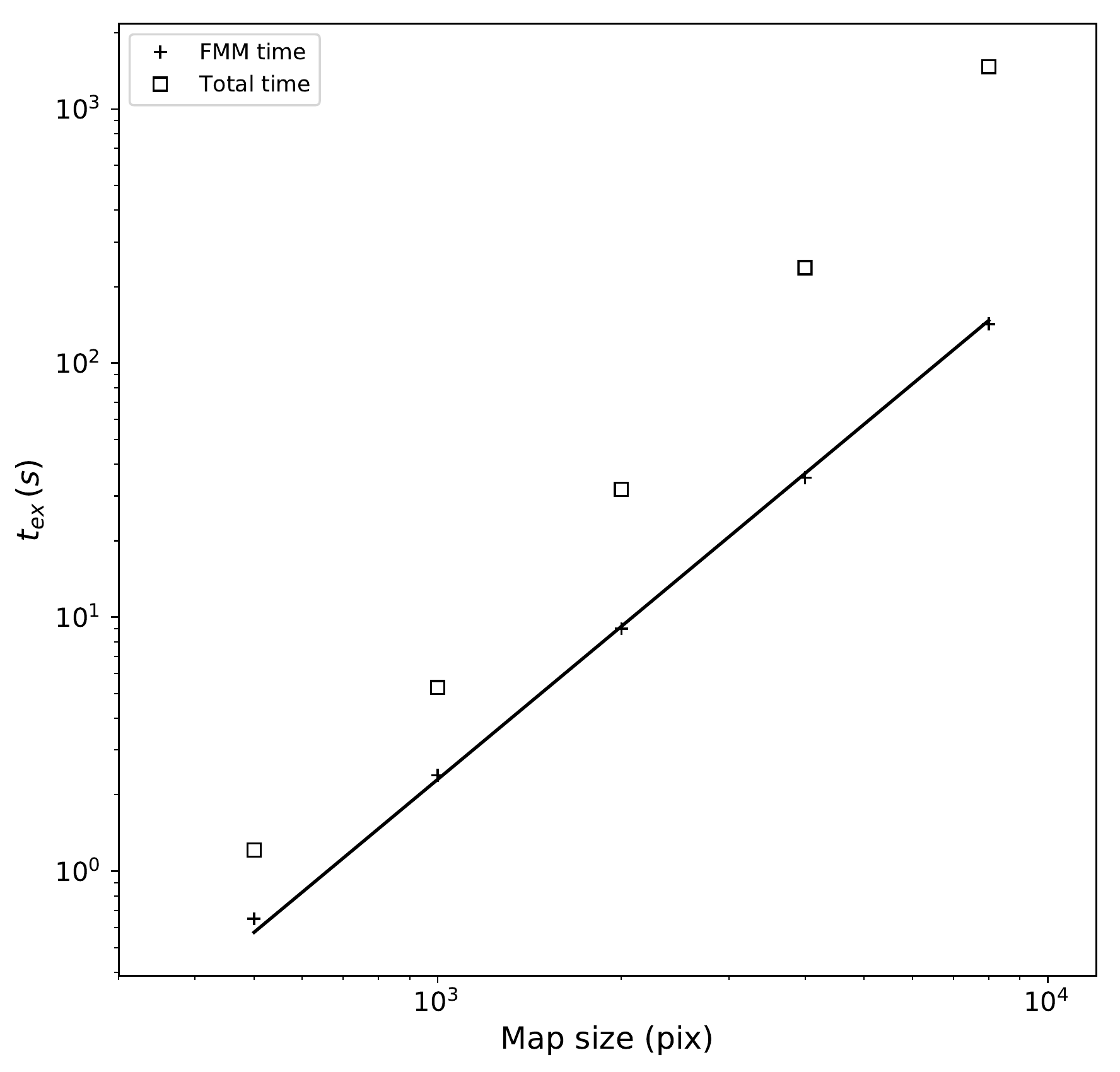}
\caption{Execution time of the FMM-IPM code vs map size.  Maps were made with $\kappa=\gamma=0.3$, with $\mu=2.5$ for a map of $L=256 E_R$ with $N_*=88682$ lenses. The line indicates linear scaling with the number of pixels in the map.\label{pwnpix}}
\end{figure}

\begin{figure}[h]
\includegraphics[width=\linewidth]{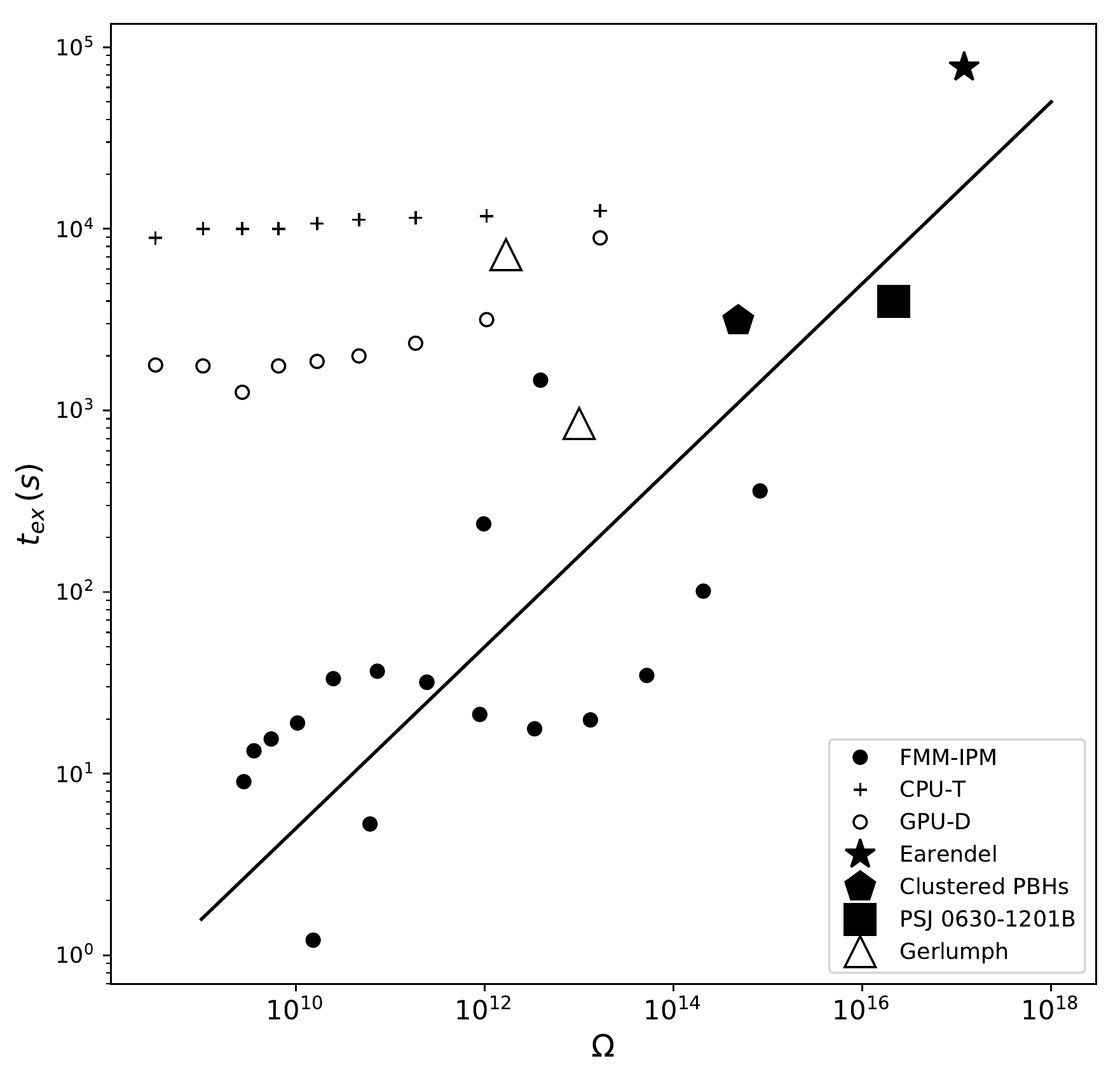}
\caption{Execution times vs workload for selected cases. Filled circles show maps with different number of deflectors and pixels. Plus symbols and open circles correspond to the CPU-T and GPU-D cases for $N_{pix}=2048^2$ in Bate et al. (2010). Open triangles show average Gerlumph maps for data releases GD0 (upper) and GD1 (lower). Pentagon shows the case of clustered PBHs. Square is the case of PSJ 0630-1201 B. Star is the case of Earendel. Straight line corresponds to $t_{ex}\propto \Omega^{1/2}$ \label{etvswl}}
\end{figure}

\begin{figure}[t]
  \includegraphics[width=\linewidth]{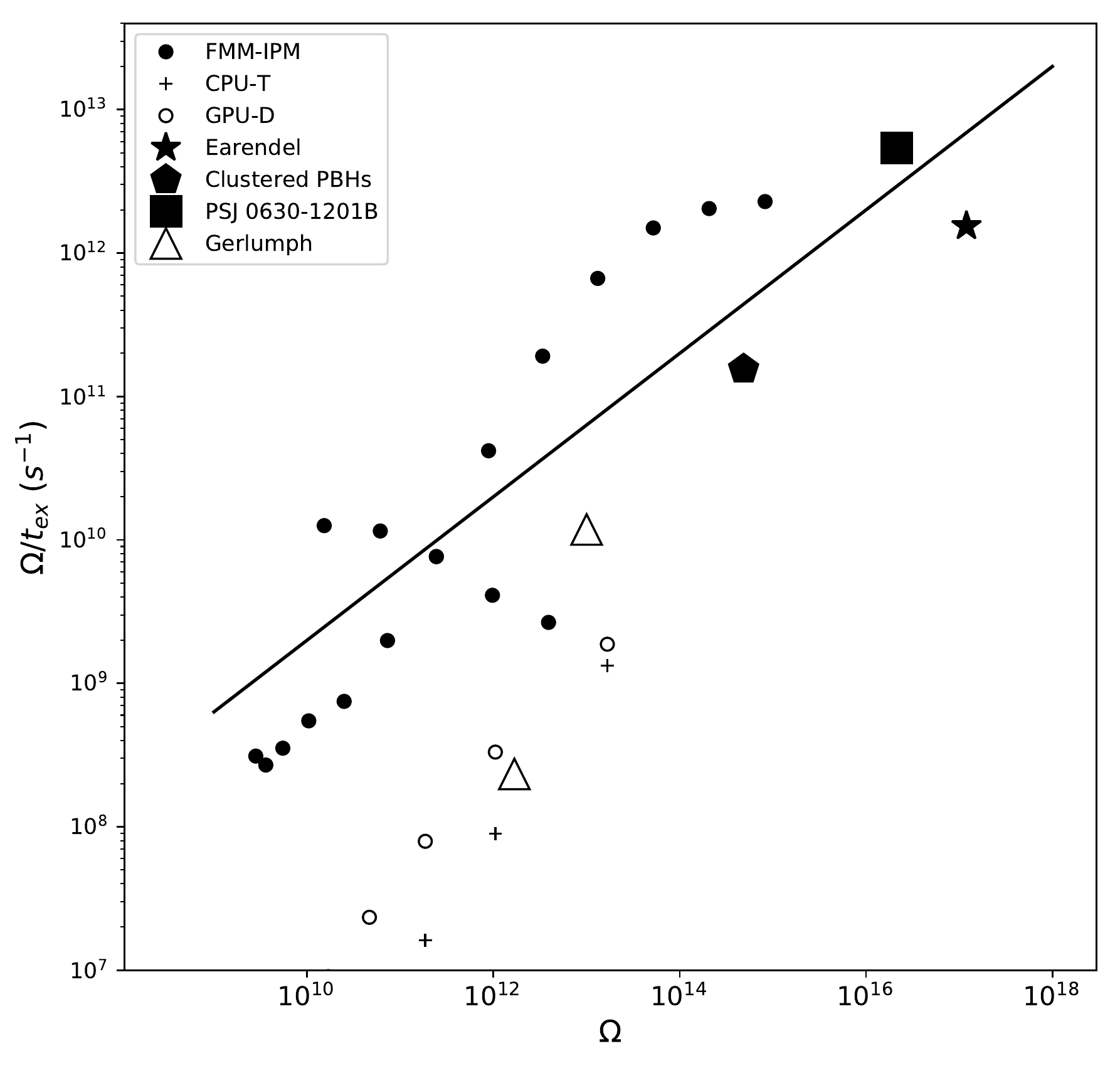}
\caption{Performance vs workload. Filled circles show maps with different number of deflectors and pixels.  Plus symbols and open circles correspond to the (best performance) CPU-T and GPU-D cases for $N_{pix}=2048^2$ in Bate et al. (2010). Open triangles show average Gerlumph maps for data releases GD0 (lower) and GD1 (upper). Pentagon shows the case of clustered PBHs. Square is the case of PSJ 0630-1201 B. Star is the case of Earendel.  Straight line corresponds to $\Omega/t_{ex}\propto \Omega^{1/2}$ \label{pvswl}}
\end{figure}

Left panel in Figure \ref{pwns} shows the execution times for a set of maps with different number of deflectors $N_*$. In order to provide comparisons with the results published with the PIP code of Shalyapin et al. (2021), we have calculated maps of the same type, with $2000\times2000$ pixels for the case $\kappa=\gamma=0.3$ with $\mu=2.5$ and 100\% of the mass in form of microlenses. We have calculated maps with size between 2 and 8192 Einstein radii, reaching a maximum of $8 \times 10^7$ lenses.
The execution time for the FMM-IPM code is rather insensitive to the number of deflectors up to $N_*\sim 10^7$, where it starts to increase linearly with $N_*$ as expected. Execution times are very similar to the ones obtained by the PIP method of Shalyapin et al. (2021), and can provide improvements of several orders of magnitudes with respect to plain IPM when a large number of deflectors is involved. 
We can also see in Figure \ref{pwns} (right) that, as expected, the best performances are reached for large number of lenses ($N_*\gtrsim 10^6$) for which the algorithm can comfortably produce performances of $\Omega/t_{ex} \sim 10^{12} s^{-1}$.

Similarly, we have made some simulations to show the scaling of execution time with the number of pixels in the map. We again performed simulations with $\kappa=\gamma=0.3$ with 100\% of mass in form of lenses, for a map of 256 Einstein radii with $N_*=88682$ lenses. Map sizes range from 500 to 8000 pixels on a side. Figure \ref{pwnpix} shows that FMM time scales linearly with $N_{pix}$ as expected.

To better show the capabilities of this new algorithm, we also find interesting to check the scaling of execution times and performances with the workload of the maps for this new code. This comparison is obviously to be taken cautiously, as it is only an approximate trend, given the strong dependence of performance with number of lenses that was mentioned above. It is nevertheless reasonably valid and interesting for maps with $N_{pix} \sim 1000^2-8000^2$ and $N_*\sim 10^3-10^8$ lenses for which the trend can provide an a priori rough estimate of the approximate execution time. The results are shown in Figures \ref{etvswl} and \ref{pvswl}. For comparison, we have included as open symbols in Figures \ref{etvswl} and \ref{pvswl} some data points from the comparison made by Bate et al. (2010) for the CPU-T and GPU-D codes for $N_{pix}=2048^2$, as well as typical Gerlumph maps for data releases GD0 and GD1.
It is particularly noticeable how the FMM-IPM on a standard personal computer can outperform the calculations of maps made with powerful GPUs.

Although the individual results show a strong dependence on the number of lenses (which is to be expected given the fact that FMM performs best for scenarios involving many lenses), there is a rough scaling $t_{ex}\sim 5\times 10^{-5} \Omega^{1/2} s$ or, equivalently, $\Omega/t_{ex}\sim 2\times 10^{4} \Omega^{1/2} s^{-1}$. Therefore, execution times for standard magnification maps of moderate magnification/resolution with $\Omega\sim 10^{12}$ are of order of one minute on a standard personal computer. High workload maps of $\Omega \sim 10^{15}-10^{16}$ can be calculated on timescales of hours. Peak performances of $\Omega/t_{ex}\sim 10^{13} s^{-1}$ can be achieved for systems with many ($\sim 10^7$) deflectors.

\section{Conclusions}
\label{conclusions}

Ray tracing is a standard procedure in numerical gravitational lensing which is needed for many purposes (e.g. finding images, calculating magnifications, time delays, etc). This procedure may be computationally expensive if many deflectors need to be considered, particularly if the number of rays to be traced is also high.
In this work we have implemented the FMM to speed up gravitational lensing ray tracing. The main results of the present work are:
\begin{itemize}
\item We incorporate to the toolbox of computational gravitational lensing a very efficient algorithm (with good a priori control on the errors), which can greatly ease and speed up this type of calculations: the Fast Multipole Method of Geengard \& Rokhlin (1987). Using this method we avoid drawbacks inherent to alternative techniques, like the simultaneous computation of deflections for a whole regular grid when using a Poisson solver (which imposes very demanding requirements on computer memory), or the re-gridding of the deflectors into the regular lattice of rays (common to mesh-particle methods) of uncertain consequences when markedly bimodal mass functions of deflectors are considered.
\item  We develop a new algorithm to calculate extragalactic microlensing magnification maps in combination with the already optimized IPM of Mediavilla et al. (2006, 2011). The new algorithm, suited even for standard personal computers or cluster nodes of modest memory capabilities, results in reductions of computing times of order $\sim 25000$ with respect to plain IRS or $\sim 500$ with respect to IPM, for large number of deflectors, making it a very competitive alternative.
\item We also show the power of the FMM algorithm for a more general lensing application by resolving the inverse lens equation (i.e. obtaining the images of the source) for a cosmological simulated lens-halo composed of $1.5\times10^7$ particles.
  
\item We demonstrate the capabilities of this new algorithm by calculating very demanding magnification maps for a small set of cases of high scientific interest that would be difficult to obtain with previous algorithms. These examples include:
  microlensing magnification maps for lensed quasars at high magnifications of large size and high resolution to allow the simultaneous study of different regions of the emitting quasar; a case of clustered Primordial Black holes, presenting an inhomogeneous distribution of lenses; and finally, a case with high resolution and extreme magnification for an individual star behind a galaxy cluster critical curve.

\item  We  present the benchmarks and performance of the new code, along with some comparisons with previous representative algorithms showing that it can achieve excelling peak performances over two orders of magnitude above previous codes even with very  modest hardware.
\item We provide a flexible and user friendly web interface for the FMM-IPM code (\url{https://gloton.ugr.es/microlensing/}), which allows researchers to easily calculate magnification maps of very large workload.
  
\end{itemize}  

\begin{acknowledgements}
   This research was financed by grants  PID2020-118687GB-C33 and PID2020-118687GB-C31, financed by MCIN/AEI/10.13039/501100011033. JJV is also financed by projects FQM-108, P20\_00334 and A-FQM-510-UGR20/FEDER, financed by Junta de Andaluc\'{\i}a. We are grateful to C. Dalla Vecchia for his help with the data from the EAGLE simulation, and to J. M. Mart\'{\i}n and M. Palomino Cobo from the IC1 for the development of the web interface. 
\end{acknowledgements}

\appendix
\section{Description of the FMM algorithm \label{AppA}}

A detailed description of the FMM algorithm is beyond the scope of this paper and can be found elsewhere (e.g. Greengard \& Rokhlin, 1987; Beatson \& Greengard, 1997; Martinsson, 2015), but we find it convenient to include here a brief general description of the main ideas behind it to help the reader to understand its peculiarities. The FMM is based on four main ideas (the two first of which are indeed shared with standard hierarchical treecodes):

\begin{enumerate}

\item A hierarchical partition of the spatial computational domain. In particular, for our 2D problem, space is progressively divided into smaller cells in a four-fold structure called a quadtree. Partitioning is carried out up to level $n\sim \log_4(N_*)$.

\item Multipole expansions (also called {\sl outer}, {\sl singular} or S expansions in the FMM jargon), which are a valid representation of the field, to the desired approximation, of a group of lenses at a certain distance.

\item Local expansions (also called {\sl inner}, {\sl regular} or R expansions), which are approximately valid representations of the field produced by groups of far away lenses, within some region.

\item Some translation theorems allow to move the center of applicability of a Multipole (M2M or S2S) or Local (L2L or R2R) expansion, and can convert a Multipole expansion into a Local one (M2L or S2R).  
    
\end{enumerate}

It is the clever introduction and extensive use of the latter two principles which allow the FMM to reduce the computational complexity with respect to standard tree-codes by reducing the number of times that loops run over all system particles. Another most important advantage of the FMM is that the error is at all time under precise control by adjusting the maximum order of the expansions $p$. Although the parameters can be optimized, as a rule of thumb, for a required accuracy $\epsilon$, expansions are truncated at order $p\sim O(-\log_2(\epsilon))$.

For the particular case of the gravitational lensing problem (which is equivalent to the electrostatic Coulombic forces in 2D), the Multipole and Local expansions have a very simple analytical form (for both, the potential and the deflection angle, or higher derivatives) when locations on the plane are expressed as complex numbers, which is extremely convenient. The deflections are therefore calculated directly from the expansions to the desired accuracy, without the need of numerical derivatives that can introduce further errors. There are also {\sl kernel independent} versions of the FMM that can also handle problems in which this expansions are not known analytically (which could, for example, account for more complicated (not point-like) {\sl pseudo-lenses} (e.g. Ying et al., 2004)).

The algorithm (as described in Greengard and Rokhlin, 1987) is therefore performed in two passes, one up and one down the tree, as follows:
\begin{enumerate}
\item Initialize:  Set refinement level to $n\sim \log_4(N_*)$ and create a quadtree structure of the plane up to that level. Set the maximum order of expansions to $p\sim \log_2(\epsilon)$.

\item Upward pass 1: Loop over boxes at the finest level, and create multipole expansions of order $p$ of the field produced by particles within that box, around the center of the box.

\item Upward pass 2: Loop over the lower levels. Use M2M translations of the multipole expansions of child boxes to build multipole expansions representing the field of all particles within coarser boxes, around their centers.  

So far, we have a description of the field produced by particles in boxes (at different levels of refinement) around their centers. Now, in the downward pass, we intend to calculate the field at a certain location by grouping particles in boxes which are not nearest neighbors into (progressively larger) boxes.
  
\item Downward pass 1: Loop over tree levels (except the finest) $l < n-1$. For each level, in an inner loop over the box's children, use M2L translations to form a local expansion about the center of each box at that level, describing the field due to all particles in the system that are not contained in the current box or its nearest neighbors. In a second inner loop over children, these local expansion are shifted, with the help of L2L translations, to the centers of the box’s children, forming the initial expansion for the boxes at the next level. Only expansion coefficients are used in this step.

\item Downward pass 2: Compute the interactions at the finest level $n$ by creating local expansions around the center of boxes at the finest levels. Use M2L translation for boxes at this level. These expansions describe the field produced by particles outside the box and nearest neighbors at the finest level. Again, only expansion coefficients are used.

We have now a local expansion of order $p$ describing the far field produced by particles outside a box and its nearest neighbors, at the finest level, and we are ready to loop over the $N_x$ locations where the field is needed.
  
\item Evaluate the local expansions of the (far) field at the required locations.

\item Add the contribution of particles in that box and the nearest neighbors directly to the far field.
  
\end{enumerate}

It is important to notice that, thanks to the extensive use of the translation theorems, loop over all particles is only performed once in the upward pass, in step 2. The rest of steps only need the coefficients of the expansions (except for the few nearby particles). This is the origin of optimized complexity of the algorithm, scaling as: $O((N_x+N_*)\times \log_2(1/\epsilon))$.


The FMM is usually assumed to be a somewhat intricate algorithm which is more difficult to code than standard hierarchical treecodes, and while this may be true, the good news is that due to its power and broad range of applicability, it is now in a very mature state, and therefore there are many available FMM libraries suitable for use {\sl out of the box} for the particular problem at hand in gravitational lensing (i.e. for the Laplace kernel) or they can be easily adapted.


\begin{thebibliography}{}
\bibitem[Barnes \& Hut(1986)]{1986Natur.324..446B} Barnes, J. \& Hut, P.\ 1986, \nat, 324, 446. doi:10.1038/324446a0
\bibitem[Bate et al.(2010)]{2010NewA...15..726B} Bate, N.~F., Fluke, C.~J., Barsdell, B.~R., et al.\ 2010, \na, 15, 726. doi:10.1016/j.newast.2010.05.008
\bibitem[Beaston \& Greengard (1997)]{BeasGreen1997} Beatson, R., Greengard, L.\ 1997: A short course on fast multipole methods. In: Wavelets, multilelvel methods and elliptic PDEs. Oxford University Press, 1–37 
\bibitem[Board \& Schulten(2000)]{2000CSE.....2a..76B} Board, J. \& Schulten, L.\ 2000, Computing in Science and Engineering, 2, 76. doi:10.1109/5992.814662
\bibitem[Cruz et al.(2011)]{2011IJNME..85..403C} Cruz, F.~A., Knepley, M.~G., \& Barba, L.~A.\ 2011, International Journal for Numerical Methods in Engineering, 85, 403. doi:10.1002/nme.2972
\bibitem[Diego et al.(2018)]{2018ApJ...857...25D} Diego, J.~M., Kaiser, N., Broadhurst, T., et al.\ 2018, \apj, 857, 25. doi:10.3847/1538-4357/aab617

\bibitem[Diego(2022)]{2022A&A...665A.127D} Diego, J.~M.\ 2022, \aap, 665, A127. doi:10.1051/0004-6361/202244027
  
\bibitem[Esteban-Guti{\'e}rrez et al.(2022a)]{2022ApJ...929L..17E} Esteban-Guti{\'e}rrez, A., Mediavilla, E., Jim{\'e}nez-Vicente, J., et al.\ 2022a, \apjl, 929, L17. doi:10.3847/2041-8213/ac57c6
\bibitem[Esteban-Guti{\'e}rrez et al.(2022b)]{2022ApJ...929..123E} Esteban-Guti{\'e}rrez, A., Ag{\"u}es-Paszkowsky, N., Mediavilla, E., et al.\ 2022b, \apj, 929, 123. doi:10.3847/1538-4357/ac57c5
 \bibitem[Fian et al.(2021)]{2021A&A...653A.109F} Fian, C., Mediavilla, E., Motta, V., et al.\ 2021, \aap, 653, A109. doi:10.1051/0004-6361/202039829
\bibitem[Garc{\'\i}a-Bellido \& Clesse(2018)]{2018PDU....19..144G} Garc{\'\i}a-Bellido, J. \& Clesse, S.\ 2018, Physics of the Dark Universe, 19, 144. doi:10.1016/j.dark.2018.01.001
  \bibitem[Garsden \& Lewis(2010)]{2010NewA...15..181G} Garsden, H. \& Lewis, G.~F.\ 2010, \na, 15, 181. doi:10.1016/j.newast.2009.06.006


\bibitem[Greengard \& Rokhlin(1987)]{1987JCoPh..73..325G} Greengard, L. \& Rokhlin, V.\ 1987, Journal of Computational Physics, 73, 325. doi:10.1016/0021-9991(87)90140-9

  \bibitem[Hilbert et al.(2009)]{2009A&A...499...31H} Hilbert, S., Hartlap, J., White, S.~D.~M., et al.\ 2009, \aap, 499, 31. doi:10.1051/0004-6361/200811054
\bibitem[Hockney \& Eastwood(1988)]{1988csup.book.....H} Hockney, R.~W. \& Eastwood, J.~W.\ 1988, Bristol: Hilger, 1988

  \bibitem[Jim{\'e}nez-Vicente et al.(2012)]{2012ApJ...751..106J} Jim{\'e}nez-Vicente, J., Mediavilla, E., Mu{\~n}oz, J.~A., et al.\ 2012, \apj, 751, 106. doi:10.1088/0004-637X/751/2/106

\bibitem[Jim{\'e}nez-Vicente et al.(2015)]{2015ApJ...799..149J} Jim{\'e}nez-Vicente, J., Mediavilla, E., Kochanek, C.~S., \& Mu{\~n}oz, J.~A.\ 2015, \apj, 799, 149 
 
\bibitem[Jim\'enez-Vicente (2016)]{2016aagl.book... J} Jim\'enez-Vicente, J.\ 2016. Tutorial on inverse ray shooting. In E. Mediavilla, J. Muñoz, F. Garzón, \& T. Mahoney (Eds.), Astrophysical Applications of Gravitational Lensing, pp. 251-290). Cambridge, UK: Cambridge University Press. doi:10.1017/CBO9781139940306.009
\bibitem[Kayser et al.(1986)]{1986A&A...166...36K} Kayser, R., Refsdal, S., \& Stabell, R.\ 1986, \aap, 166, 36
\bibitem[Kelly et al.(2018)]{2018NatAs...2..334K} Kelly, P.~L., Diego, J.~M., Rodney, S., et al.\ 2018, Nature Astronomy, 2, 334. doi:10.1038/s41550-018-0430-3
\bibitem[Kochanek(2004)]{2004ApJ...605...58K} Kochanek, C.~S.\ 2004, \apj, 605, 58. doi:10.1086/382180

\bibitem[Martinsson (2015)]{2015Marttinsson} Martinsson, PG. \ 2015. Fast Multipole Methods. In: Engquist, B. (eds) Encyclopedia of Applied and Computational Mathematics. Springer, Berlin, Heidelberg. https://doi.org/10.1007/978-3-540-70529-1\textunderscore448

  
\bibitem[Mediavilla et al.(2006)]{2006ApJ...653..942M} Mediavilla, E., Mu{\~n}oz, J.~A., Lopez, P., et al.\ 2006, \apj, 653, 942

\bibitem[Mediavilla et al.(2011)]{2011ApJ...741...42M} Mediavilla, E., Mediavilla, T., Mu{\~n}oz, J.~A., et al.\ 2011, \apj, 741, 42
  \bibitem[Mediavilla et al.(2015)]{2015ApJ...814L..26M} Mediavilla, E., Jim{\'e}nez-Vicente, J., Mu{\~n}oz, J.~A., et al.\ 2015, \apjl, 814, L26. doi:10.1088/2041-8205/814/2/L26

  \bibitem[Mediavilla et al.(2016)]{2016aagl.book.....M} Mediavilla, E., Mu{\~n}oz, J.~A., Garz{\'o}n, F., et al.\ 2016, Astrophysical Applications of Gravitational Lensing, by Evencio Mediavilla, Jose A. Mu{\~n}oz, Francisco Garz{\'o}n, Terence J. Mahoney, Cambridge, UK: Cambridge University Press, 2016

\bibitem[Meena et al.(2022)]{2022MNRAS.514.2545M} Meena, A.~K., Arad, O., \& Zitrin, A.\ 2022, \mnras, 514, 2545. doi:10.1093/mnras/stac1511

\bibitem[Metcalf \& Petkova(2014)]{2014MNRAS.445.1942M} Metcalf, R.~B. \& Petkova, M.\ 2014, \mnras, 445, 1942. doi:10.1093/mnras/stu1859

\bibitem[Morgan et al.(2010)]{2010ApJ...712.1129M} Morgan, C.~W., Kochanek, C.~S., Morgan, N.~D., et al.\ 2010, \apj, 712, 1129. doi:10.1088/0004-637X/712/2/1129
\bibitem[Petkova et al.(2014)]{2014MNRAS.445.1954P} Petkova, M., Metcalf, R.~B., \& Giocoli, C.\ 2014, \mnras, 445, 1954. doi:10.1093/mnras/stu1860  

  \bibitem[Popovi{\'c} et al.(2020)]{2020A&A...634A..27P} Popovi{\'c}, L. {\v{C}}., Afanasiev, V.~L., Moiseev, A., et al.\ 2020, \aap, 634, A27. doi:10.1051/0004-6361/201936088

  \bibitem[Potter et al.(2017)]{2017ComAC...4....2P} Potter, D., Stadel, J., \& Teyssier, R.\ 2017, Computational Astrophysics and Cosmology, 4, 2. doi:10.1186/s40668-017-0021-1

\bibitem[Schaye et al.(2015)]{2015MNRAS.446..521S} Schaye, J., Crain, R.~A., Bower, R.~G., et al.\ 2015, \mnras, 446, 521. doi:10.1093/mnras/stu2058


\bibitem[Schneider \& Weiss(1987)]{1987A&A...171...49S} Schneider, J. \& Weiss A.\ 1987, \aap, 171, 49

\bibitem[Shajib et al.(2019)]{2019MNRAS.483.5649S} Shajib, A.~J., Birrer, S., Treu, T., et al.\ 2019, \mnras, 483, 5649. doi:10.1093/mnras/sty3397

\bibitem[Shalyapin et al.(2021)]{2021A&A...653A.121S} Shalyapin, V.~N., Gil-Merino, R., \& Goicoechea, L.~J.\ 2021, \aap, 653, A121. doi:10.1051/0004-6361/202140527
\bibitem[Thompson et al.(2010)]{2010NewA...15...16T} Thompson, A.~C., Fluke, C.~J., Barnes, D.~G., et al.\ 2010, \na, 15, 16. doi:10.1016/j.newast.2009.05.010

\bibitem[Venumadhav et al.(2017)]{2017ApJ...850...49V} Venumadhav, T., Dai, L., \& Miralda-Escud{\'e}, J.\ 2017, \apj, 850, 49. doi:10.3847/1538-4357/aa9575
  
\bibitem[Vernardos \& Fluke(2013)]{2013MNRAS.434..832V} Vernardos, G. \& Fluke, C.~J.\ 2013, \mnras, 434, 832. doi:10.1093/mnras/stt1076
\bibitem[Vernardos et al.(2014)]{2014ApJS..211...16V} Vernardos, G., Fluke, C.~J., Bate, N.~F., et al.\ 2014, \apjs, 211, 16. doi:10.1088/0067-0049/211/1/16

  
\bibitem[Wambsganss(1999)]{1999JCoAM.109..353W} Wambsganss, J.\ 1999, Journal of Computational and Applied Mathematics, 109, 353

\bibitem[Wambsganss(2006)]{2006glsw.conf..453W} Wambsganss, J.\ 2006, Saas-Fee Advanced
 Course 33: Gravitational Lensing: Strong, Weak and Micro, 453

 \bibitem[\protect\citeauthoryear{Wang, Yokota, \& Barba}{2021}]{2021JOSS....6.3145W} Wang T., Yokota R., Barba L., 2021, JOSS, 6, 3145. doi:10.21105/joss.03145
\bibitem[Welch et al.(2022)]{2022Natur.603..815W} Welch, B., Coe, D., Diego, J.~M., et al.\ 2022, \nat, 603, 815. doi:10.1038/s41586-022-04449-y

\bibitem[Ying et al.(2004)]{2004JCoPh.196..591Y} Ying, L., Biros, G., \& Zorin, D.\ 2004, Journal of Computational Physics, 196, 591. doi:10.1016/j.jcp.2003.11.021
\bibitem[Zheng et al.(2022)]{2022ApJ...931..114Z} Zheng, W., Chen, X., Li, G., et al.\ 2022, \apj, 931, 114. doi:10.3847/1538-4357/ac68ea

\end{thebibliography}
\end{document}